\documentclass{article}
\usepackage{dcase2023,amsmath,amsfonts,graphicx,url,times,booktabs, tabularx}
\usepackage{float}
\usepackage{xcolor}
\usepackage{pifont}
\usepackage{multirow}

\newcommand{\norm}[1]{\left\lVert#1\right\rVert}
\newcommand{\cmark}{\ding{51}}%
\newcommand{\xmark}{\ding{55}}%

\usepackage{xcolor}
\usepackage{cite}

\definecolor{orange_m}{HTML}{FF7F0E}
\definecolor{blue_m}{HTML}{1F77B4}

\title{Advancing Natural-Language Based Audio Retrieval \\ with PaSST and Large Audio-Caption Data Sets}

\name{Paul Primus$^1$, Khaled Koutini$^{1,2}$, Gerhard Widmer$^{1,2}$}
\address{
$^1$Institute of Computational Perception (CP-JKU)  \\
$^2$LIT Artificial Intelligence Lab\\
Johannes Kepler University, Austria
}

\begin{document}

\ninept
\maketitle

\begin{sloppy}

\begin{abstract}
This work presents a text-to-audio-retrieval system based on pre-trained text and spectrogram transformers. Our method projects recordings and textual descriptions into a shared audio-caption space in which related examples from different modalities are close. Through a systematic analysis, we examine how each component of the system influences retrieval performance. As a result, we identify two key components that play a crucial role in driving performance: the self-attention-based audio encoder for audio embedding and the utilization of additional human-generated and synthetic data sets during pre-training. We further experimented with augmenting ClothoV2 captions with available keywords to increase their variety; however, this only led to marginal improvements. Our system ranked first in the 2023's DCASE Challenge, and it outperforms the current state of the art on the ClothoV2 benchmark by 5.6 pp. mAP@10.
\end{abstract}

\begin{keywords}
Natural-Language-Based Audio Retrieval, PaSST, ChatGPT 
\end{keywords}

\section{Introduction}

\label{sec:intro}

Natural-language-based audio retrieval revolves around ranking audio recordings based on their relatedness to textual descriptions. Current state-of-the-art methods in this domain are based on the dual-encoder approach which converts both recordings and textual descriptions into high-level representations and then aligns them within a shared audio-caption space. The ranking of candidate audios is carried out by measuring their distance to the textual descriptions in the shared embedding space. The dual-encoder setup has been widely adopted in audio retrieval systems \cite{Xie2022_results,xu2022_t6b,mei2022_t6b,pellegrini2022_t6b}, because it permits fast ranking and the use of pre-trained audio and text embedding models. CNN architectures pre-trained on AudioSet~\cite{panns} are the most common audio encoders and large transformer models, such as BERT~\cite{bert} and RoBERTa~\cite{roberta}, are the most popular text encoders. Recently, Mei et al. \cite{wavecaps} managed to set a new state-of-the-art performance on ClothoV2~\cite{clotho} by introducing WavCaps \cite{wavecaps}, a large dataset with synthetic captions. 

In this work, we elaborate on our findings in the context of subtask 6b of the 2023 DCASE challenge, which is concerned with natural-language-based audio retrieval. Our method is also based on the dual-encoder method but differs from previous methods in three central aspects:
firstly, we use an audio spectrogram transformer called PaSST \cite{passt} instead of CNN14 \cite{panns} for audio embedding because PaSST outperforms CNN14 on AudioSet~\cite{audioset} and other relevant audio benchmarks \cite{hear}. PaSST employs Patchout \cite{passt} during training, which increases training speed and memory efficiency while at the same time acting as a regularizer. Swapping CNN14 for PaSST improves the retrieval performance significantly, as we will show in Section~\ref{sec:audio_embedding}.
Secondly, we pre-train our models on AudioCaps \cite{audiocaps} and WavCaps \cite{wavecaps}, two large audio captioning datasets with human-generated and synthetic captions, to alleviate the data scarcity problem. Using these large datasets to pre-train the models results in significantly better retrieval performance, as we will show in Section~\ref{sec:data_set}.
And thirdly, we take advantage of the available metadata associated with each audio recording and augmented the captions during training by generating additional captions using the available keywords and the GPT3.5-turbo API, also known as ChatGPT. This augmentation method successfully reduces overfitting during fine-tuning but only results in a minor performance improvement (see Section \ref{sec:fine}). Our implementation and the keyword-augmented captions are available in our GitHub repository\footnote{\url{https://github.com/OptimusPrimus/dcase2023_task6b}}.

\section{Method}

\begin{table*}[ht]
\begin{center}
\begin{tabular}{@{}l|l@{}}
\toprule
Instruction & GPT3.5-turbo's Response                                                                                                                                                                                                                                                                                                                                                                                                           \\ \midrule
\begin{tabular}[c]{@{}l@{}}I will give a description of an audio recording and tags associated with the \\ audio recording. Generate 5 audio captions describing the sound event.\\ Each audio caption should be one sentence with less than 20 words. Use \\ grammatical subject-verb-object sentences. Do not include names of cities,\\ countries, and persons. Do not include the time. Do not include recording \\ device names. Do not write introductions or explanations. Do not use \\ “heard” or “recorded”. Start each output sentence with its index. \\ \textcolor{blue_m}{'A person is making noise by tapping their fingernails on a solid surface.'} \\ {[}\textcolor{orange_m}{dull, metal}{]}\end{tabular} & \begin{tabular}[c]{@{}l@{}}1. The sound of nails tapping on a \textcolor{orange_m}{metal} surface is heard.\\ 2. Fingernails are producing a \textcolor{orange_m}{dull} sound when tapped.\\ 3. A person is tapping their nails on a solid, \textcolor{orange_m}{metal} surface.\\ 4. The noise of someone tapping their fingernails is audible.\\ 5. Nails are rhythmically tapping on a hard \textcolor{orange_m}{metal} object.\end{tabular} \\ \bottomrule
\end{tabular}
\end{center}

\caption{Example query fed to GPT3.5-turbo to augment a ClothoV2 caption (\textcolor{blue_m}{in blue}) with the available keywords (\textcolor{orange_m}{in orange}; query inspired by WavCaps \cite{wavecaps}). The response is a list of rephrased captions, some of which take the keywords into account (highlighted \textcolor{orange}{in orange}).}
\label{table:query_result}
\end{table*}

Our model uses separate audio and caption embedding networks, denoted as $\phi_a(\cdot)$ and $\phi_c(\cdot)$, respectively, to embed pairs of spectrograms and descriptions $\{(a_i, c_i)\}_{i=1}^{N}$ into a shared $D$-dimensional space such that representations of matching audio-caption pairs are close. This behavior is achieved by contrastive training, which makes the embeddings of matching audio-caption pairs $(a_i, c_i)$ more similar while pushing the representations of mismatching pairs $(a_i, c_{j; j \neq i})$ apart. The agreement between audio $a_i$ and description $c_j$ is estimated via the normalized dot product in the shared embedding space:
$$C_{ij} = \frac{\phi_{\textrm{a}}(a_{i})^T \cdot \phi_{\textrm{c}}(c_{j})}{\norm{\phi_{\textrm{a}}(a_{i})}^2 \norm{\phi_{\textrm{t}}(c_{j})}^2}$$
The similarity matrix $\mathbf{C} \in \mathbb{R}^{N \times N}$ holds the agreement of matching pairs on the diagonal and the agreement of mismatching pairs off-diagonal. We train the system using the NT-Xent \cite{NTxent} loss, which is defined as the Cross-Entropy ($\mathrm{CE}$) between the ground truth and the posterior over the text queries and the audio recordings; the ground truth is given by the identity matrix $\mathbf{I} \in \mathbb{R}^{N \times N}$:
$$\mathcal{L} = \frac{1}{2\cdot N} \sum_{i=1}^{N} \textrm{CE}(\mathbf{C}_{i*}, \mathbf{I}_{i*}) + \textrm{CE}(\mathbf{C}_{*i}, \mathbf{I}_{*i})$$

\subsection{Audio Embedding Models}
\label{sec:pass_setup}
We choose the Patchout faSt Spectrogram Transformer (PaSST) \cite{passt} to convert audio recordings into a compact, high-level vector representation because it achieves state-of-the-art results on multiple audio classification benchmarks \cite{Koutini22PaSSTHear} while keeping memory and computational complexity low compared to the vanilla audio spectrogram transformers~\cite{ast}. PaSST uses ImageNet \cite{imagenet} pre-trained parameters from a vision transformer \cite{vit,deit} and fine-tunes them on AudioSet \cite{audioset} for general-purpose audio tagging. The relatively low computational and memory footprint is achieved by dropping patches from the input sequence. This procedure, called Patchout~\cite{passt}, additionally regularizes the model during training. We conducted experiments with PaSST models that take audios of up to ten seconds in length as input and extract overlapping or non-overlapping patches of size $16 \times 16$ from the input spectrogram. Pre-trained PaSST models are available on GitHub\footnote{\url{https://github.com/kkoutini/passt_hear21}}. We additionally experimented with two convolutional neural networks pre-trained on AudioSet, namely CNN10 and CNN14 \cite{panns}. These models can handle inputs of arbitrary length, so we directly input up to 30 seconds long audio instead of cutting them into shorter segments. Table \ref{tab:passt} gives an overview of all audio embedding models used in our experiments.

\begin{table}[H]
\centering
\begin{tabular}{@{}lllclr@{}}
\toprule
          & \begin{tabular}[c]{@{}l@{}}patch\\ stride\end{tabular} & \begin{tabular}[c]{@{}l@{}}patch\\ out\end{tabular} & \begin{tabular}[c]{@{}l@{}}input\\ length (s)\end{tabular} & \begin{tabular}[c]{@{}l@{}}AS\\ mAP\end{tabular} & \begin{tabular}[c]{@{}l@{}}number\\ params\end{tabular} \\ \midrule
CNN10     & -                                                      & -                                                   & 30                                                     & $38.0$                                                 & 6.3M                                                        \\
CNN14     & -                                                      & -                                                   & 30                                                     & $43.1$                                                 & 81.8M                                                       \\ \midrule
PaSST-L   & $10 \times 10$                                         & $4; 50$                                       & 10                                                     & $45.9$                                                 & 41.8M                                                       \\
PaSST-N   & $16 \times 16$                                         & $2; 15$                                       & 10                                                     & $46.8$                                                 & 86.2M                                                       \\
PaSST-S   & $10 \times 10$                                         & $4; 50$                                       & 10                                                     & $48.6$                                                 & 86.2M                                                       \\
PaSST-S20 & $10 \times 10$                                         & $4; 80$                                       & 20                                                     & $47.4$                                                 & 86.2M                                                       \\ \bottomrule
\end{tabular}
\caption{Overview of the audio embedding models compared in our experiments. The first section shows CNNs from \cite{panns}. The second section summarizes PaSST variants\cite{passt}.}
\label{tab:passt}
\end{table}

\begin{table*}[ht]

\centering
    \scalebox{0.92}{
\begin{tabular}{@{}llcclccllllc@{}}
\toprule
& \begin{tabular}[x]{@{}c@{}}audio\\embedding\end{tabular}  & \begin{tabular}[x]{@{}c@{}}segment\\length (s)\end{tabular} & overlap    & \begin{tabular}[x]{@{}c@{}}text\\embedding\end{tabular} & finetune & \begin{tabular}[x]{@{}c@{}}GPT-\\augment\end{tabular}   & mAP@10 & R@1   & R@5   & R@10 & \\ \midrule
& PaSST-N & 10 & \xmark  & bert-small & \xmark & \xmark    & 32.98 & 21.45 & 48.71 & 62.05 &\\
& PaSST-N & 10 & \xmark  & bert-base & \xmark & \xmark    & 35.22 & 23.07 & 51.48 & 65.36 & \\
& PaSST-N & 10 & \xmark  & bert-large & \xmark & \xmark    & 35.78 & 23.75 & 52.27 & 65.57 & \\
& PaSST-N & 10 & \xmark & roberta-base & \xmark & \xmark  & 35.12 & 23.02 & 51.89 & 65.26 & \\
& PaSST-N & 10 & \xmark   & roberta-large & \xmark & \xmark & 36.65 & 24.26 & 53.89 & 66.87 & \\ \midrule
& CNN10  & 30 & \xmark   & bert-base & \xmark & \xmark     & 23.72 & 14.18 & 36.59 & 49.21 & \\
& CNN14  & 30 & \xmark   & bert-base & \xmark & \xmark    & 28.06 & 17.86 & 40.82 & 54.56 & \\
& PaSST-L  & 10 & \cmark   & bert-base & \xmark & \xmark     & 33.47 & 21.67 & 49.24 & 63.16 & \\
& PaSST-N  & 10 & \xmark   & bert-base & \xmark & \xmark     & 35.22 & 23.07 & 51.48 & 65.36 &\\
& PaSST-S  & 10 & \cmark  & bert-base & \xmark & \xmark  & 32.83 & 20.90 & 48.82 & 62.60 & \\ \midrule
& PaSST-N & 10 & \xmark    & roberta-large  & \cmark & \xmark  & 38.00 & 25.51 & 55.06 & 68.56 &  \\
& PaSST-N & 10 & \xmark  & roberta-large  & \cmark & \cmark& 38.56 & 26.07 & 55.27 & 69.30 &\\ \midrule
\cite{dcase2023_task6b} & CNN14 & 30 & -  & all-mpnet-base-v2  & \xmark & \xmark & 22.20  & 13.00 & 34.30 & 48.00 & \\
\cite{wavecaps} & CNN14 & 30 & -  & bert-base  & \cmark & \xmark & 32.95  & 21.41 & 47.77 & 62.10 & \\ 

\bottomrule
\end{tabular}
}
\caption{Text-to-audio retrieval performance on the ClothoV2 test set for different combinations of language and audio embedding models (sections one and two, respectively). The impact of additional fine-tuning on ClothoV2 and ClothoV2GPT is shown in section three. Section four shows results from the DCASE baseline system \cite{dcase2023_task6b} and the current state of the art \cite{wavecaps} (values based on WavCaps's GitHub repository).}
\label{table:results}

\end{table*}

\subsection{Sentence Embedding Models}
We compared five different sentence embedding models: bert-small, bert-base, bert-large, roberta-base, and roberta-large. All models are bi-directional self-attention-based sentence encoders that underwent self-supervised pretraining on the BookCorpus \cite{bookcorpus} and WikiText datasets \cite{wikitext}. BERT- and RoBERTa-based models differ in the masking strategy used during training: the former was trained using 10 fixed masks for each sentence, while the latter used new, dynamically generated masks in each forward pass. For both models, we selected the output vector that corresponds to the class token as sentence embedding. The parameter counts for bert-small, bert-base, bert-large, roberta-base, and roberta-large are around 29, 110, 345, 123, and 354 million, respectively.

\subsection{Shared Audio-Caption Space}

The audio and text embeddings generated by the encoders are integrated into a shared audio-caption space by using a simple linear projection that maps the embedding models' output to a size of $1024$. Initial experiments suggested that using a non-linear projection (e.g., a multilayer neural network) does not significantly improve performance.

\subsection{Datasets}

Our final models were trained in two steps on multiple datasets. First, we performed pretraining on ClothoV2, AudioCaps, and WavCaps. The resulting models were then further finetuned on a custom, augmented version of ClothoV2 (called ClothoV2\_GPT below), which also takes into account the available meta-data associated with each audio file.

\subsubsection{ClothoV2}

ClothoV2 \cite{clotho} contains $10$-$30$ second-long audio recordings and captions that are between 8 and 20 words long. The development set's training, validation, and test split suggested by the organizers contains 3840, 1045, and 1045 recordings, respectively, and each recording is associated with five human-generated captions. The leaderboard evaluation split used for the final system ranking contains 1000 audio recordings and 1000 captions. We used the validation split to monitor the generalization performance and report the performance on the test split in Section \ref{section:results}.

\subsubsection{AudioCaps}
AudioCaps \cite{audiocaps} contains $51,308$ audio recordings taken from AudioSet and one human-written caption for each of them. Each audio recording has a duration of 10 seconds, and the captions are, on average, 9.8 words long. We concatenated the training, validation, and testing split of AudioCaps into one large dataset and used it for pretraining.

\subsubsection{WavCaps}
WavCaps \cite{wavecaps} is a weakly-labeled audio-caption dataset that contains $403,050$ audio recordings of varying length collected from FreeSound, BBC Sound Effects, SoundBible, and a strongly labeled subset of AudioSet \cite{audioset_strong}. Each audio file is associated with a synthetic audio caption that was created by instructing the GPT3.5-turbo model to extract relevant sound events from metadata and output a single-sentence description. The generated captions are, on average, 7.8 words long. The authors demonstrated the usefulness of these synthetic captions by successfully using the dataset for audio retrieval, audio captioning, and text-based sound generation.

\subsubsection{ClothoV2\_GPT}

Each audio recording in the ClothoV2 dataset is associated with additional metadata consisting of the file name, a list of keywords, a username, and a web URL. We took advantage of the additional information and used GPT3.5-turbo to augment the human-generated captions with the available keywords. To this end, we instructed the model to take the keywords into account and rephrase the available captions. We generated five new captions for each caption in the training set, resulting in $96,000$ additional captions. Table \ref{table:query_result} gives an example query and the resulting augmented captions: GPT3.5-turbo successfully combined the description and the keywords into five varying descriptions with similar content; four of these contain the provided keywords. We will refer to this augmented version of ClothoV2 as ClothoV2\_GPT. The generated captions are available in our GitHub repository.

\subsection{Preprocessing}
\label{sec:preprocess}
To allow batched processing of recordings of varying lengths, we extracted random 30-second snippets from those audio recordings that are longer than 30 seconds and zero-padded shorter recordings to the maximum duration in the current batch. The resulting waveforms were converted to $128$-bin log-MEL spectrograms using a $1024$-point FFT ($32$ms) and hop size of $320$ ($10$ms). The spectrograms were centered and whitened with the approximate global mean and standard deviation before feeding them into the audio embedding model.
The input sentences were pre-processed by transforming all characters to lowercase and removing punctuation. The resulting strings were tokenized with the WordPiece tokenizer, padded to the maximum sequence length in the current batch, and truncated if they were longer than 32 tokens.

\subsection{Training}
\label{sec:train}
We pre-trained the models on AudioCaps, WavCaps, and the training set of ClothoV2. Both embedding models were jointly optimized using gradient descent with a batch size of 64. We used the Adam update rule \cite{adam} for 16 epochs, with one warmup epoch. Thereafter, the learning rate was reduced  from $2 \times 10^{-5}$ to $10^{-7}$ using a cosine schedule. The hyperparameters of the optimizer were set to PyTorch's \cite{pytorch} defaults. We further used structured patchout as a regularizer during training with hyperparameters depending on the audio length and patch extraction (see Table~\ref{tab:passt}).  Finetuning on ClothoV2\_GPT was done in a similar manner as pretraining but only for five epochs with a maximum learning rate of $8 \times 10^{-6}$. During the finetuning procedure, a caption was swapped with one of its five GPT-augmented versions with a probability of $0.3$. 

\section{Results}
\label{section:results}
The performance of different audio and sentence embedding models is summarized in Table \ref{table:results}. The combination of PaSST-N and roberta-large, pre-training on all data sets, and fine-tuning with GPT augmentation outperforms the current state of the art \cite{wavecaps} based on bert-base and CNN14 by $5.6$ pp. mAP@10. In the following sections, we analyze our method in detail to identify each component's impact on the overall performance.

\subsection{Pre-training Data Sets}
\label{sec:data_set}
ClothoV2 is relatively small compared to captioning data sets in the image domain; to further enhance the performance, we additionally leveraged AudioCaps and WavCaps. In this section, we investigate the impact of the additional pre-training data sets on the final performance. To this end, we used PaSST-N and bert-base and trained them on different combinations of the three sets. We report the results in terms of mAP@10 on ClothoV2's test set in Table \ref{tab:pre-train}. Pretraining on WavCaps or ClothoV2 results in similar performance of around 27 mAP@10; pretraining exclusively on AudioCaps is roughly 6 pp.~worse.  Surprisingly, adding AudioCaps to WavCaps did not further improve the result; however, adding ClothoV2 to WavCaps or AudioCaps yielded improvements of 6.8 and 9.6 pp., respectively. The best result overall was achieved by combining all three data sets.

\begin{table}[H]
\centering
\scalebox{0.9}{
\begin{tabular}{@{}lllc@{}}
\toprule
AudioCaps     & WavCaps     & ClothoV2     & mAP@10 \\ \midrule
\cmark & \xmark & \xmark & 21.01  \\
\xmark & \cmark & \xmark & 27.62  \\
\xmark & \xmark & \cmark & 27.28  \\ \midrule
\cmark & \cmark & \xmark & 27.13  \\
\xmark & \cmark & \cmark & 34.42  \\
\cmark & \xmark & \cmark & 30.64  \\ \midrule
\cmark & \cmark & \cmark & \textbf{35.22}  \\ \bottomrule
\end{tabular}
}

\caption{Ablation study on the effect of pre-training data sets.}
\label{tab:pre-train}
\end{table}

\subsection{Text Embedding Models}

We assumed that larger sentence embedding models would lead to better retrieval performance. To test this hypothesis, we experiment with three variants of BERT and two variants of RoBERTa. The results are summarized in the first section of Table \ref{table:results}. Larger BERT sentence encoders indeed performed better (compare bert-small, bert-base, and bert-large), and a similar trend can be observed for RoBERTa (compare roberta-base and roberta-large). The best overall results were achieved by utilizing roberta-large.

\subsection{Audio Embedding Models}
\label{sec:audio_embedding}

We likewise assumed that using a self-attention-based architecture would lead to further improvements. To test this assumption, we compared two convolutional architectures (CNN10 and CNN14) to three recent spectrogram tansformers (PaSST-L, PaSST-N, and PaSST-S); section two of Table \ref{table:results} summarizes the results. Scaling up the audio embedding model from CNN10 to CNN14 yielded an improvement of $4.3$ pp.~mAP@10. Switching from CNN14 to PaSST-N further improved the mAP@10 by $7.1$ pp. PaSST-S, which extracts overlapping spectrogram patches and performs better on AudioSet, surprisingly did not further improve the retrieval performance over PaSST-N. This inconsistency could be due to a suboptimal patchout configuration.

\subsection{Audio Context Length}
PaSST uses a learnable positional encoding with a fixed length of 10 seconds and consequently cannot handle longer audio segments. To deal with the up to 30-second long audio recordings in ClothoV2, we cut longer waveforms into shorter segments, embedded each segment separately, and averaged the resulting embeddings over time to obtain a single vector representation. To investigate the impact of the segment length, we conducted experiments by splitting the recordings into 2, 5, 10, and 15 seconds long snippets. We used PaSST-S20 for those experiments, an architecture similar to PaSST-S, but with a positional encoding for audios of up to 20 seconds in length. The results are given in Figure \ref{fig:segment_length}. While a longer context is advantageous performance-wise, it also comes at the price of higher computational cost (which grows quadratically with the input size). We find that PaSST's default maximum input length of ten seconds strikes a good balance.

\begin{figure}[h]
    \centering
    \includegraphics[width=0.8\linewidth]{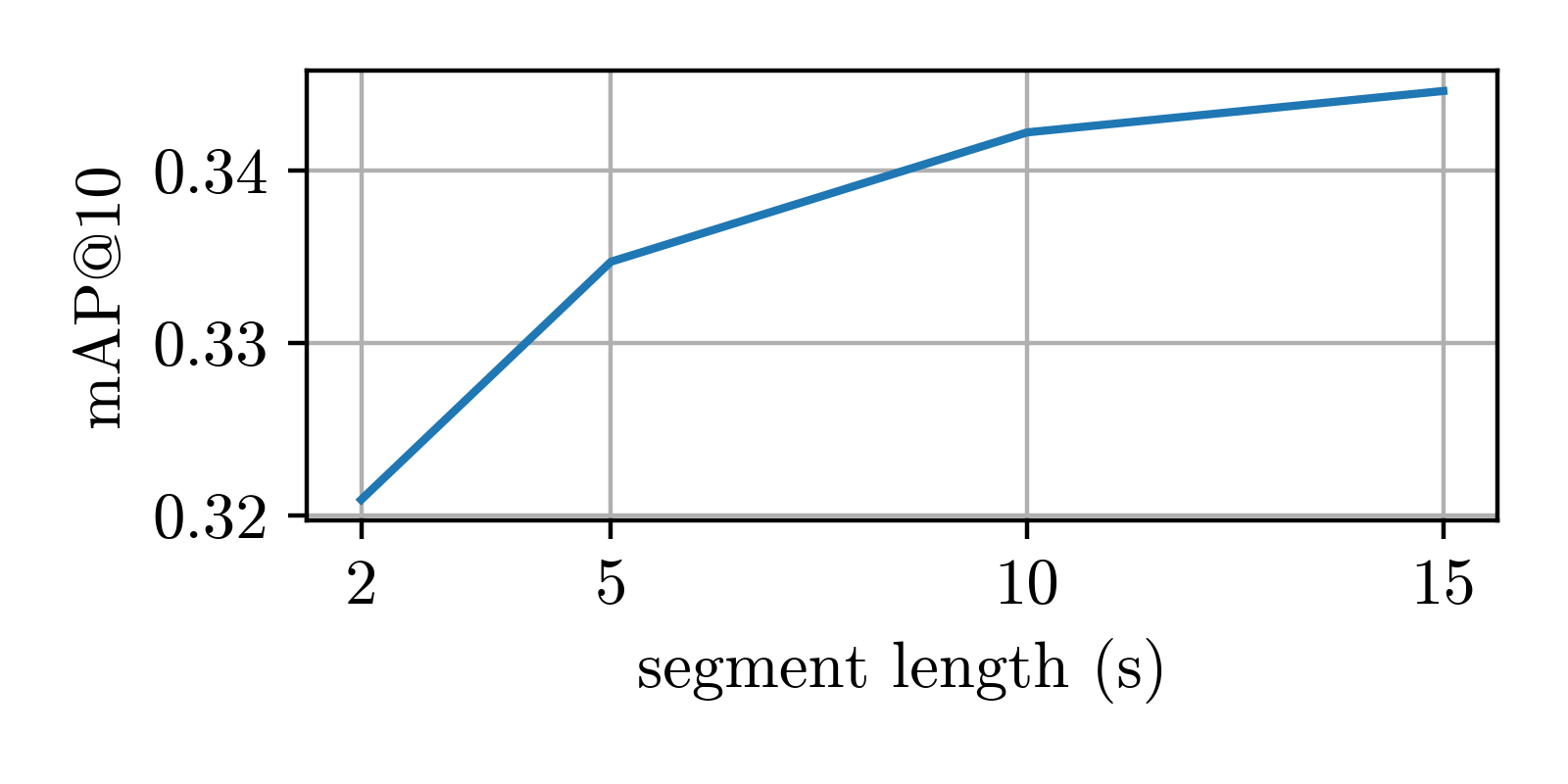}
    \caption{Performance of PaSST-S20 for different audio lengths.}
    \label{fig:segment_length}
\end{figure}

\subsection{Fine-Tuning}
\label{sec:fine}

We further investigated the impact of additional finetuning on the target data sets. To this end, we continued training (as described in Section \ref{sec:train}) on the training split of ClothoV2 with and without GPT augmentation. The results are given in the third section of Table \ref{table:results}. Finetuning on ClothoV2 without GPT augmentation further improved the pre-trained model by 1.3 pp. mAP@10. When finetuned with GPT-Augmentation, overfitting was slightly reduced, and the model improved by 1.9 pp. mAP@10. A similar advantage for the GPT-augmented data set can be observed for the hidden test set of the 2023's DCASE Challenge.

\section{Discussion \& Conclusion}

This work presented a dual-encoder system for automatic audio retrieval, achieving state-of-the-art results on the ColthoV2 benchmark. The results of our experiments attribute the considerable performance gains to two factors: firstly, the additional data sets with human-generated and synthetic captions, and secondly, the audio spectrogram transformer, which scaled better with the additional data compared to convolutional neural networks. Augmentation of the captions with the additional keywords reduced overfitting during finetuning; however, it did not significantly improve retrieval performance. One possible explanation for this is that each recording in the ClothoV2 training set is associated with five different captions, which are likely to contain the most relevant keywords already; adding further captions increases the variety only marginally.


\section{ACKNOWLEDGMENT}
\label{sec:ack}
The LIT AI Lab is financed by the Federal State of Upper Austria. The computational results presented in this work have been partially achieved using the Vienna Scientific Cluster (VSC).

\bibliographystyle{IEEEtran}
\bibliography{template}
%
%
%
%
%
%
%
%
%

\end{sloppy}
\end{document}